\documentstyle[pra,aps,twocolumn]{revtex}
\begin{document}
\title{Atomic detection and matter-waves coherence}
\author{E. V. Goldstein and Pierre Meystre\\
Optical Sciences Center, University of Arizona, Tucson, AZ 85721}
\maketitle
\begin{abstract}
We analyze several models of atomic detectors in the context of
the measurement of coherence properties of matter waves. In particular,
we show that an ionization scheme measures normally-ordered correlation
functions of the Schr\"odinger field, in analogy with the optical situation.
However, it exhibits a sensitivity to exchange processes that is normally
absent in optics.
\end{abstract}
%\twocolumn
Optical coherence theory is based on the observation that most quantum
measurements that can be performed on the electromagnetic field yield a
signal proportional to normally ordered correlation functions of that
field \cite{Gla65}. A quantized multimode field
is then said to be coherent to order $N$ if all normally ordered correlation
functions up to order $N$ factorize.
No such theory is presently available for atomic coherence,
probably because until recently it had not been necessary to think of atomic
samples as Schr\"odinger fields. But the experimental work on
ultracold atoms, BEC
\cite{AndEnsMat95,DavMewAnd95,EnsJinMat96,MewAndDru96,BraSacHul97}
and atom lasers \cite{MewAndKur97} has changed that situation,
and the need for a proper theory of atomic coherence is now quite urgent
\cite{KetMie97}.

At least for the case of bosonic fields, it is tempting to simply transpose
Glauber's coherence theory \cite{Gla65}.
This approach has been the de facto situation
so far, but appealing as it might sound, it must be applied with
caution, due to the fundamental difference between electromagnetic and
matter-wave fields. Most optical experiments detect light by absorption,
i.e. by ``removing'' photons from the light field. This is the
reason why normally ordered correlation functions are so important. But
atomic detectors work in a number of different ways: One can chose to
measure electronic properties, or center-of-mass properties, or both.
While one detector might be sensitive to atomic velocities, another
might measure local densities and a third electronic properties
only. Additional difficulties arise from the fact that atomic fields are
self-interacting, which significantly complicates the propagation of
atomic coherence as compared to the case of light. From these remarks, it
should be clear that a theory of matter waves coherence is much richer than
its optical equivalent. Yet, like Glauber's coherence theory, it
should be operational and based on explicit detection schemes.

The goal of this note is to analyze several ideal atom detectors and to
determine which correlation functions of the matter-wave field they are
sensitive to. The systems we explicitly consider are a nonresonant
atomic imaging system such as used e.g. in the MIT BEC experiments, and
a detector working via atomic ionization. We show that while
the off-resonance imaging detector is sensitive to density correlation
functions, a narrow-band ionization detector measures normally ordered
correlation functions of the Schr\"odinger field itself, in analogy with the
optical case. Intermediate situations are more complicated, due to the
quadratic dispersion of matter waves. Higher-order detection schemes
also involve exchange terms usually absent in the optical case. \\

{\em Nonresonant imaging}

Consider first atomic detection by non-resonant imaging, a situation where
a strongly detuned electromagnetic field interacts with the atoms in the
sample in such a way that it induces only virtual transitions. We consider
for concreteness ground state atoms described by the Schr\"odinger field
operator ${\hat \Psi}({\bf r})$ with
$[{\hat \Psi}({\bf r}), {\hat \Psi}^\dagger({\bf r'})] =
\delta({\bf r} -{\bf r}')$
for bosons, and decompose the electromagnetic field into a classically
populated mode of wave vector ${\bf k}_0$ and polarization
$\bbox {\epsilon}_0$ and a series of weakly excited sidemodes of
wave vectors ${\bf k}_\ell$ and polarizations ${\bbox \epsilon}_\ell$.
After adiabatic elimination of the upper electronic state of the atomic
transition under consideration, the interaction between the Schr\"odinger
field and the radiation field is described to lowest order in the
side-modes by the effective Hamiltonian
\begin{eqnarray}
V & = & \hbar \int d^3r
\frac{|\Omega_0({\bf r})|^2}{\delta_0}
{\hat \Psi}^\dagger ({\bf r})
{\hat \Psi} ({\bf r})
\nonumber\\
&+&\hbar\sum_\ell \int d^3 r
\left (\frac
{\Omega_0({\bf r}) \Omega_\ell^\star}{\delta_0}a_\ell^\dagger
e^{i({\bf k}_0-{\bf k}_\ell)\cdot {\bf r}}\right .
\nonumber\\
&+&\left.
\frac{\Omega_0^\star({\bf r}) \Omega_\ell}{\delta_0} a_\ell
e^{-i({\bf k}_0-{\bf k}_\ell)\cdot {\bf r}}
\right){\hat \Psi}^\dagger ({\bf r}) {\hat \Psi} ({\bf r}),
\end{eqnarray}
where ${\bf k}_\ell$ is the
wave vector of the $\ell$-th mode of the field, of frequency $\omega_\ell$
and polarization ${\bbox \epsilon}_\ell$, the sum is over all
field modes in the quantization volume $V$, and ${\cal E}_\ell =
[\hbar \omega_\ell/2\epsilon_0 V]^{1/2}$ is the ``electric field per
photon'' of mode $\ell$. The annihilation and creation operators $a_\ell$
and $a^\dagger_\ell$ satisfy the boson commutation relation $[a_\ell,
a^\dagger_{\ell '}] = \delta_{\ell, \ell'}$. We have also introduced the
Rabi frequencies $\Omega_0({\bf r})=d {\cal E}_0({\bf r})
({\bbox \epsilon}\cdot{\bbox\epsilon}_0)/\hbar$
and $\Omega_\ell=d {\cal E}_\ell
({\bbox \epsilon}\cdot{\bbox \epsilon}_\ell)/\hbar$, and
the atom-field detuning $\delta_0\equiv \omega_a-\omega_0$ is assumed
to be much larger than $\Omega_0$, $\delta_0\gg \Omega_0({\bf r})$.

Assuming that the electromagnetic field is initially in the state
$|{\cal E} \rangle $ and the Schr\"odinger field in the state $|\phi_g
\rangle $, the probability that the system undergoes a transition
from that to another state is given to first order in perturbation
theory by
\begin{eqnarray}
w &=& \frac{1}{\delta_0^2}  \sum_{\ell,\ell '} \Omega_\ell^\star
\Omega_{\ell '}\int d^3r \Omega_0({\bf r})\int d^3 r'
\Omega_0^\star({\bf r}')
\nonumber\\
& &\int_0^{\Delta t} dt \int_0^{\Delta t} dt'
\langle \phi_g |{\hat \rho}({\bf r}, t){\hat \rho}({\bf r}', t')|\phi_g
\rangle
\nonumber \\
&\times&\langle {\cal E}|a_\ell^\dagger a_{\ell '} e^{i({\bf k}_{\ell '}
\cdot{\bf r}'-{\bf k}_\ell\cdot{\bf r})}
e^{-i(\omega_{\ell '}t'-\omega_\ell t)}+h.c.  |{\cal E} \rangle
\end{eqnarray}
where the Schr\"odinger wave density is defined as
\begin{equation}
{\hat\rho}({\bf r},t) \equiv {\hat \Psi}^\dagger({\bf r},t)
{\hat \Psi}({\bf r},t)
\end{equation}
and $\Psi({\bf r},t)=U^\dagger\Psi({\bf r})U$ is the time-dependent
Schr\"odinger field in the interaction representation with respect to the
atomic Hamiltonian, i.e. $U=\exp(-i {\cal H}_At/\hbar)$.

We further assume for concreteness that all electromagnetic sidemodes are
initially in a vacuum. The measurement on the Schr\"odinger
field is then carried out by detecting photons scattered by the atoms into
the sidemodes, in a fashion familiar from resonance fluorescence
experiments. The most important non-trivial contribution to the fluorescence
signal is proportional to the intensity $|\Omega_0|^2$ of the incident
field,
\begin{eqnarray}
w &=& \frac {|\Omega_0|^2}{\delta_0^2}
\sum_\ell|\Omega_\ell|^2 \int d^3r  \int d^3 r'
\int_0^{\Delta t} dt \int_0^{\Delta t} dt'
\nonumber \\
& & e^{i({\bf k}_0 -{\bf k}_\ell)\cdot ({\bf r}-{\bf r}') }
e^{-i(\omega_0 -\omega_\ell)(t-t')}
\nonumber\\
&\times&
\langle \phi_g |{\hat \rho}({\bf r}, t){\hat \rho}({\bf r}', t')|\phi_g
\rangle ,
\end{eqnarray}
and hence is sensitive to the second-order correlation function of the
sample density. This is to be compared to the results of Javanainen
\cite{Jav95},
who showed that the the spectrum of the scattered radiation
is a function of
$\langle {\hat \rho} ({\bf r},0){\hat \rho}({\bf r}, t)\rangle$.
Indeed, it can be shown in all generality that any measurement involving
the electromagnetic field scattered by the atomic sample under conditions
of off-resonant imaging are determined by correlation functions of the
Schr\"odinger field density.\\

{\em Ionization}

The reason off-resonant imaging yields a signal dependent on ${\hat \rho}
({\bf r}, t)$ is that the electric dipole interaction is bilinear in the
Schr\"odinger field operators. This difficulty can however be eliminated if,
instead of making measurements on the radiation field, one detects the
atoms directly. One scheme that achieves this goal is the ionization
method that we now discuss.

Consider a detector consisting of a tightly focussed laser
beam that can ionize atoms by inducing transitions from their ground
electronic level $|g \rangle $ to a continuum level $| i \rangle $.
\footnote {Hot wire detectors can be modeled in a similar fashion.}
The corresponding single-particle Hamiltonian is
\begin{equation}
H = H_{cm} + H_{el} + V({\bf r})  \equiv H_0 + V
\end{equation}
where $H_{cm}$ is the center-of-mass Hamiltonian, $H_{el}$ the
electronic Hamiltonian, and $V({\bf r})$ describes the electric dipole
interaction between the atom and the ionizing laser field. $H_{el}$ has
eigenstates $\varphi_n$ and eigenfrequencies $\omega_n$,
$H_{el} |\varphi_n \rangle = \hbar \omega_n |\varphi_n \rangle$.
The corresponding atomic manybody Hamiltonian is
\begin{equation}
{\cal H}_0 = \int d^3 r{\hat \Psi}^\dagger({\bf r}) H_0{\hat \Psi}({\bf r})
\end{equation}
where in the Born-Oppenheimer approximation ${\hat \Psi}({\bf r})$ is a
multicomponent field with components ${\hat \Psi}_n({\bf r})$.

We are interested in measuring properties of the ground state component
${\hat \Psi}_g({\bf r})$ of this field, which is dipole-coupled to continuum
states ${\hat \Psi}_i({\bf r})$. We assume for simplicity that the
center-of-mass wave function of these latter states is well described by
plane waves of momentum ${\bf q}$, so that ${\cal H}$ may be expressed
as
\begin{equation}
{\cal H}_0 = {\cal H}_g + \sum_i {\cal H}_i ,
\end{equation}
where
\begin{equation}
{\cal H}_i = \hbar\sum_{\bf q} (\omega_q + \omega_i) b_{{\bf q},i}^\dagger
b_{{\bf q},i}.
\end{equation}
Here we expanded ${\hat \Psi}_i({\bf r})$ in plane waves as
${\hat \Psi}_i({\bf r}) = \sum_{\bf q} \phi_{i,{\bf q}}({\bf r})
b_{{\bf q},i}$ with $[b_{{\bf q},i}, b^\dagger_{{\bf q}',i'}] =
\delta_{{\bf q}{\bf q}'}\delta_{ii'}$, and $\omega_q = \hbar q^2/2M$.
(Note that the inclusion of ground state collisions is straightforward
and does not affect our conclusions.)

In terms of the components ${\hat \Psi}_n({\bf r})$ of the Schr\"odinger
field, the electric dipole interaction Hamiltonian is
\begin{equation}
{\cal V} = \hbar \sum_i \int d^3 r \Omega_i({\bf r})
{\hat \Psi}_i^\dagger ({\bf r}){\hat \Psi}_g({\bf r}) + H.c.,
\label{vion}
\end{equation}
where $\Omega_i$ is the Rabi frequency between the
levels $|g \rangle $ and $|i \rangle $, and the laser field is
treated classically.

In this detection scheme, one extracts
information about the state of the field ${\hat \Psi}_g({\bf r}, t)$ by
measuring, e.g. the number of atoms in the continuum. For atoms cooled well
below the recoil temperature and tightly focused laser beams, the spatial
size of the atomic wave function is much larger than the laser spot and we
can approximate the electric field ${\bf E}({\bf r})$ by ${\bf E}({\bf r})
\simeq {\bf E} \delta({\bf r}-{\bf r}_0)$, so that Eq. (\ref{vion}) becomes
\begin{equation}
{\cal V} = \hbar \sum_i \Omega_i({\bf r}_0)
{\hat \Psi}_i^\dagger ({\bf r_0}){\hat \Psi}_g({\bf r_0}) + H.c.
\end{equation}

We take the atomic system to be initially in the state
\begin{equation}
|\psi(0)  \rangle = |\{\psi_{i,{\bf q}}(0)\}, \psi_g(0) \rangle .
\end{equation}
To first order in perturbation theory, the transition probability away
from that state during the time interval $\Delta t$ is
\begin{eqnarray}
w & = & \sum_{i,q} |\Omega_i({\bf r}_0)|^2
\int_0^{\Delta t} dt \int_0 ^{\Delta t} dt'
\nonumber\\
& &\langle \psi_{i,{\bf q}}(0) |{\hat \Psi}_i({\bf r}_0, t)
{\hat \Psi}_i^\dagger({\bf r}_0,t') |\psi_{i,{\bf q}}(0) \rangle
\nonumber\\
&\times&
\langle \psi_g(0) |{\hat \Psi}_g^\dagger({\bf r}_0, t)
{\hat \Psi}_g ({\bf r}_0,t') |\psi_g (0)\rangle + c.c.
\label{w}
\end{eqnarray}
There is a fundamental distinction between the present situation and
Glauber's photodetection theory, because in the present case both
the detected and detector fields consist of
matter waves. There is a complete symmetry between these two fields so far,
and their roles are interchangeable.
In order to break this symmetry and to truly construct a detector,
we now make a series of assumptions on the state of the detector fields
${\hat \Psi}_i({\bf r}, t)$. Physically, this amounts to making a statement
about the way the detector is prepared prior to a measurement. Specifically,
we assume that all atoms are in the ground state, $\Psi_i({\bf r}_0)
|\psi_{i, {\bf q}}(0) \rangle =|0 \rangle $, and that any atom in an
ionized state will be removed from the sample
instantaneously. In that case, the second term in Eq. (\ref{w}) vanishes
and we have
\begin{eqnarray}
w &=& \sum_i |\Omega_i({\bf r}_0)|^2
\int_0^{\Delta t} dt \int_0 ^{\Delta t} dt'
\nonumber\\
& &\sum_q e^{i  \omega_q(t-t')} \phi_q({\bf r}_0) \phi^\star_q({\bf r}_0)
\nonumber \\
&\times&
e^{i \omega_i(t-t')}\langle \psi_g |{\hat \Psi}_g^\dagger({\bf r}_0, t)
{\hat \Psi}_g ({\bf r}_0,t') |\psi_g \rangle .
\label{wd}
\end{eqnarray}

At this point, it is convenient to distinguish three different operating
regimes: In the first one, only one final electronic state is considered,
and in addition a velocity selector is used to filter just those ionized
atoms with a given center-of-mass momentum. We call this a {\em
narrowband single-state detector.} The second scheme allows for a broader
velocity filter, but still considers a single continuum electronic state,
and we call it a {\em broadband single-state detector}. Finally, we also
discuss a {\em general broadband detector} where neither the final momentum
state nor the final electronic state is narrowly selected.

More precisely, a narrowband single-state detector includes a velocity
selector with a bandwidth $\Delta {\bf q}$ around a central value
${\bf q}_0$ such that for the detection times $\Delta t$ of interest,
one has $\Delta t \Delta \omega_{\bf q} \ll 1$, where $\Delta
\omega_{\bf q} = \hbar {\bf q}_0\Delta {\bf q}/2M$. In this case and for 
a stationary Schr\"odinger fields Eq.(\ref{wd}) reduces to
\begin{eqnarray}
r_{nb}(\omega,\omega_{{\bf q}_0})&=&
\frac{\Delta \omega_{\bf q}^3}{c^3}|\Omega({\bf r}_0)|^2
\nonumber\\
&\times&
 \int_0^{\Delta t} d\tau e^{-i(\omega+\omega_{{\bf q}_0}) \tau}
G_{A}(0,\tau;{\bf r}_0,{\bf r}_0),
\end{eqnarray}
where we dropped the index $i$ of the observed continuum state
for clarity, introduced the ionization rate $r_{nb}(\omega,\omega_{\bf q})
= w_{nb}(\omega,\omega_{\bf q})/ \Delta t$ and defined
the atomic normally ordered first-order ground state
correlation function
$$G_{A}(t,t';{\bf r}_0,{\bf r}_0)=
\langle \phi_g |{\hat \Psi}_g^\dagger({\bf r}_0, t)
{\hat \Psi}_g ({\bf r}_0,t') |\phi_g \rangle. $$
From the Wiener-Khintchine theorem, we recognize that for
large enough $\Delta t$, the detector measures the
spectrum of the Schr\"odinger field ${\hat \Psi}_g({\bf r}_0, 0)$.

In the case of a broad single-state detector, in contrast, we have
\begin{eqnarray}
r_{1b} &\simeq&|\Omega({\bf r}_0)|^2\int_0^{\Delta t}d\tau e^{-i\omega \tau}
G_{pr}(0,\tau;{\bf r}_0,{\bf r}_0)
\nonumber\\
&\times&G_{A}(0,\tau;{\bf r}_0,{\bf r}_0)
\end{eqnarray}
where we have introduced the center-of-mass propagator
\begin{equation}
G_{pr}(t_1,t_2;{\bf r}_1,{\bf r}_2)
=\sum_q \phi_q({\bf r}_1)\phi_q^\star({\bf r}_2)e^{i\omega_q(t_2-t_1)}.
\end{equation}
In that case, the ionization rate is proportional to the Fourier transform
of the product of $G_{pr}(0,\tau;{\bf r}_0,{\bf r}_0)$ and the
correlation function $G_{A}(0,\tau;{\bf r}_0,{\bf r}_0)$, or in other words
to the convolution of the Fourier transforms of these functions. The Fourier
transform of the center-of-mass propagator can therefore be interpreted as
the spectral resolution of the detector.

We finally turn to the case of a general broad-band detector, where the
spectrum of the detector is much broader than the spectrum of
the detected quantity.  Assuming that the spectrum of the atomic correlation
function is centered at $\bar \omega$, we find
\begin{equation}
r_{bb} \simeq \eta({\bf r}_0) G_{A}(0,0;{\bf r}_0,{\bf r}_0),
\end{equation}
where we have introduced the ``detector efficiency''
\begin{equation}
\eta({\bf r}_0)=\int d\tau \sum_i |\Omega_i({\bf r}_0)|^2
\langle\Psi_i({\bf r}_0,\tau)\Psi^\dagger_i({\bf r}_0,0)\rangle
e^{-i\bar{\omega} \tau} .
\label{eta}
\end{equation}
As expected, a broadband detector is not able to resolve any spectral
feature of the Schr\"odinger field, and only measures the local atomic
density. \\

{\em Higher-order correlations}

The detection of higher-order correlation functions of the Schr\"odinger
field can be achieved by a straightforward generalization of the ionization
detector. For instance, second-order coherence measurements can be carried
out by focussing the laser at two locations ${\bf r}_1$ and ${\bf r}_2$, in
which case
\begin{equation}
{\cal V} = \hbar\sum_{\mu=1,2} \sum_i \Omega_i({\bf r}_\mu)
{\hat \Psi}^\dagger_i({\bf r}_\mu) {\hat \Psi}_g({\bf r}_\mu) + H.c.
\end{equation}
Assuming as before that the continuum states are initially empty
and for a general broadband detector, the joint probability to
ionize an atom at ${\bf r}_1$ and another one at ${\bf r}_2$ is then
\begin{eqnarray}
w_2 &\simeq& \eta({\bf r}_1,{\bf r}_2)\eta({\bf r}_2,{\bf r}_1)
\int_0^{\Delta t}dt_1\int_0^{\Delta t}dt_2
\nonumber\\
& &\langle {\hat \Psi}_g^\dagger({\bf r}_1, t_1){\hat \Psi}_g^\dagger
({\bf r}_2, t_2){\hat \Psi}_g({\bf r}_2, t_1)
{\hat \Psi}_g({\bf r}_1, t_2) \rangle
\nonumber \\
&+& \eta({\bf r}_1)\eta({\bf r}_2)
\int_0^{\Delta t}dt_1\int_0^{\Delta t}dt_2
\nonumber\\
& &\langle {\hat \Psi}_g^\dagger({\bf r}_1, t_1){\hat \Psi}_g^\dagger
({\bf r}_2, t_2){\hat \Psi}_g({\bf r}_2, t_2)
{\hat \Psi}_g({\bf r}_1, t_1) \rangle ,
\label{final}
\end{eqnarray}
where we have introduced the detector cross efficiency
$\eta({\bf r}_1, {\bf r}_2)$ as a straightforward generalization of Eq.
(\ref{eta}). The first term in Eq. (\ref{final}) is an exchange term
resulting from the fact that the detector field  is a single Schr\"odinger
field. It results from the interference between detectors at points
${\bf r}_1$ and ${\bf r}_2$.
The second term is the usual term also appearing in the double
photo-detection of optical fields. In that latter case, the exchange
term does not appear because the two detectors used to measure the
field are taken to be distinguishable. Note also that in the position
measurement scheme proposed in Ref. \cite{ThoWan94}, interferences
do not occur as the set of states ionized at each location are taken
to be distinguishable. We finally remark that as a consequence of the
exchange term, the signal cannot simply be expressed in terms of
correlation functions of ${\hat \rho}({\bf r}, t).$

In summary, we have analyzed several detectors that permit to access
different classes of correlation functions of the Schr\"odinger field. Most
interesting perhaps is the ionization scheme, which is closely related to
the detectors familiar from the detection of optical fields. However, it
presents new features, and is in particular sensitive to exchange
processes. But ionization detectors make destructive measurements. This is
in contrast to off-resonant imaging, which is nondestructive but measures
density correlation functions instead of the more familiar normally-ordered
correlation functions of the Schr\"odinger field itself.

\acknowledgements
This work is supported in part by the U.S. Office of Naval Research
Contract No. 14-91-J1205, by the National Science Foundation Grant
PHY95-07639, by the U.S. Army Research Office and by the
Joint Services Optics Program.

%\bibliography{quasi_new}

\end{document}